\documentclass[11pt]{article}
\usepackage{amsmath}
\usepackage{amsfonts}
\usepackage{graphicx}
\usepackage{geometry}
\usepackage{setspace}

\title{Navigating Inflation in Ghana: How Can Machine Learning Enhance Economic Stability and Growth Strategies}
\author{
    Theophilus G. Baidoo\textsuperscript{1}, Ashley Obeng\textsuperscript{2} \\
    \\
    \textsuperscript{1}Department of Epidemiology and Biostatistics, \\
    Indiana University - Bloomington. \\
    \textsuperscript{2} Department of Mathematical Sciences, \\
    University of Texas at El Paso. \\
}
\date{}

\begin{document}

\maketitle

\begin{abstract}
Inflation remains a persistent challenge for many African countries. This research investigates the critical role of machine learning (ML) in understanding and managing inflation in Ghana, emphasizing its significance for the country's economic stability and growth. Utilizing a comprehensive dataset spanning from 2010 to 2022, the study aims to employ advanced ML models, particularly those adept in time series forecasting, to predict future inflation trends. The methodology is designed to provide accurate and reliable inflation forecasts, offering valuable insights for policymakers and advocating for a shift towards data-driven approaches in economic decision-making. This study aims to significantly advance the academic field of economic analysis by applying machine learning (ML) and offering practical guidance for integrating advanced technological tools into economic governance, ultimately demonstrating ML's potential to enhance Ghana's economic resilience and support sustainable development through effective inflation management.
\end{abstract}


\newpage
\section{Introduction}
The global economic landscape has recently encountered significant challenges, culminating
in a state of deadlock. This predicament arises from various factors, including unstable
economic growth, fluctuating commodity prices, diminished consumer spending, and the
uneven distribution of resources. These issues have collectively deteriorated the standard of living worldwide, plunging many into financial distress. Central to these economic upheavals is inflation, a phenomenon that not only affects Ghana but also has a global impact. Inflation signifies the rate at which the general level of prices for goods and services rises eroding money's purchasing power. It is a critical macroeconomic indicator, and managing it is vital for maintaining economic stability. Nations endeavor to curb its adverse effects through effective monetary policies. Ghana's historical struggle with inflation, particularly since the mid-1960s, exemplifies its significance. The country has witnessed periods of high inflation,
driven by considerable budgetary deficits and substantial monetary expansion. Since 2008, the Bank of Ghana has aimed for single-digit inflation to foster investment and encourage sustained economic growth. The advent of the COVID-19 pandemic has aggravated these pre-existing challenges, causing unprecedented disruptions in economic activities and amplifying inflationary pressures. Ghana, alongside other countries, has been significantly affected by the pandemic, which has exacerbated economic hardships and heightened inflation's impact on the nation and its populace. The pandemic has profoundly influenced the global economy, disrupting supply chains, altering consumer behavior, and necessitating changes in government policies. This has introduced a new layer to the efforts to manage and comprehend inflation within the framework of Ghana's economic evolution, underscoring the complex interplay between global events and national economic stability.

\section{Review of Literature}
This section of the literature review delves into the various studies and theories aimed at unraveling the causes behind inflation, a critical economic phenomenon with global
repercussions. A thorough comprehension of these causes is imperative for devising effective policy measures. The study by Appiah \cite{appiah2011forecasting} analyzes exchange rate fluctuations between the Ghana Cedi and the US Dollar from 1994 to 2010. Utilizing the ARIMA model, the research identifies the ARIMA (1, 1, 1) configuration as the optimal fit, highlighting a consistent depreciation trend of the Ghana Cedi relative to the US Dollar. Abdul-Karim \cite{iddrisu2019modeling} shifts the focus towards the volatility characteristics of Ghana's inflation rates spanning the years 2000 to 2018. Through the application of ARCH, GARCH, and EGARCH models, it was determined that the EGARCH (12, 1) model most accurately represents the data, demonstrating an upward trajectory in the overall price levels of goods and services. Further exploring regional inflation dynamics, Okyere \cite{okyere2015econometric} delves into the modeling and forecasting of inflation rates within the Volta Region, utilizing ARIMA techniques over the period from January 2009 to September 2015. The findings underscore the ARIMA (2, 1, 2)
model's efficacy in predicting inflation, which consistently hovered in the double-digit range. In a quest to refine short-term inflation forecasting for Ghana, Maurice \cite{omane2013determining} examines data from January 1971 to October 2012, comparing Seasonal-ARIMA and Holt-Winters methodologies. The study concludes that the Seasonal-ARIMA (ARIMA (2, 1, 1) (0, 0, 1)12) framework stands out for its accuracy, as evidenced by superior metrics (MAE, RMSE, MAPE, MASE) in comparison to the Holt-Winters approach.
Expanding the analytical scope, \cite{nortey2014modelling} evaluate the performance of ARCH, GARCH, and EGARCH models against monthly inflation data from January 2000 to December 2013. The combined EGARCH (1, 2) model with an ARIMA (3, 1, 2) mean equation emerged as the most fitting, adeptly capturing the complexities of Ghana's inflationary trends. Senanu \cite{klutse2020inflation} investigates the potential of SARIMA models for inflation forecasting, emphasizing their capacity to effectively account for seasonal fluctuations. This exploration highlights the significant advantage of SARIMA models in encapsulating the cyclical patterns inherent in inflationary movements.

\section{Methodology}
\subsection{Source of Data}
This study utilized monthly inflation rate data from the Bank of Ghana and the Ghana Statistical Service. The analysis specifically concentrated on the period from 2010 to 2022, offering a focused overview of inflation trends during these years. While the analysis primarily focused on the period from 2010 to 2022, data from 2022 was specifically reserved for forecasting purposes for the time series model, enabling a focused examination of inflation trends and predictive modeling for future rates.

\subsection{Trend analysis of Inflation rates}

\begin{figure}[h!]
 \centering
  \includegraphics[width=10cm]{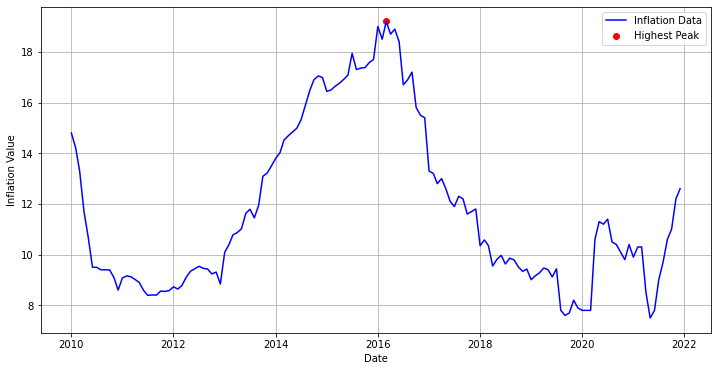}\\
  \caption{Time series plot of Inflation in Ghana from 2010 to 2021}\label{look2}
\end{figure}

The plot of inflation in Ghana from 2010 to 2021 in Figure \ref{look2} reveals a dynamic economic landscape, characterized by significant fluctuations and periods of stabilization. The most striking feature is the peak in March 2016, when the inflation rate soared to 19.6\%, indicating a period of considerable economic instability or transformation. Despite this overall downward trend, there were occasional spikes in inflation, though these were less severe compared to the earlier years. Such intermittent increases could be attributed to external economic shocks or temporary fiscal challenges. 

\subsection{Stationarity}
\begin{figure}[h!]
 \centering
  \includegraphics[width=10cm]{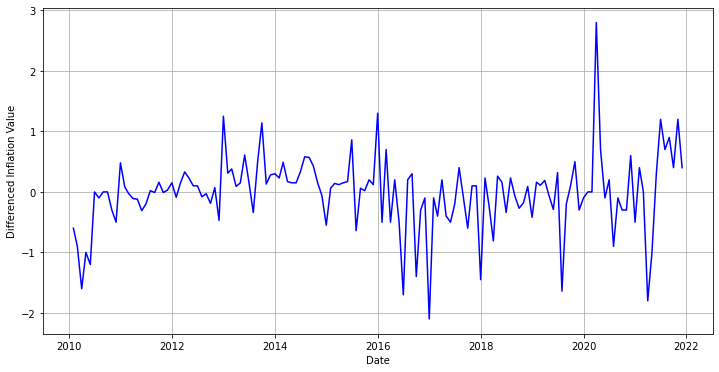}\\
  \caption{Time series plot of Inflation in Ghana from 2010 to 2021 after differencing}\label{look3}
\end{figure}

In assessing the inflation rates from 2010 to 2021, it was crucial to examine the stationarity of the time series data, as the presence of a unit root can significantly affect analysis and interpretations. Two widely recognized tests were employed: the Augmented Dickey-Fuller (ADF) test \cite{dickey1979distribution} and the Kwiatkowski-Phillips-Schmidt-Shin (KPSS) test \cite{kwiatkowski1992testing}. The ADF test yielded a test statistic of -1.63 and a p-value of 0.46. Since the test statistic did not fall below the critical values (1\%: -3.48, 5\%: -2.88, 10\%: -2.57) and the p-value exceeded the conventional threshold of 0.05, the null hypothesis of a unit root could not be rejected, indicating that the inflation rate time series may not be stationary. To address this non-stationarity, first differencing was applied, which involves calculating the difference between consecutive observations, thereby focusing on changes in the inflation rate. After differencing, the ADF test showed significant improvement with a test statistic of -9.62 and a p-value less than 0.05, strongly rejecting the null hypothesis of a unit root. The test statistic was well below the critical values, confirming the stationarity of the differenced series. Additionally, the KPSS test on the differenced data yielded a test statistic of 0.1584 and a p-value of 0.1, indicating that the test statistic remained below the critical values and further supporting the absence of a unit root. These findings validate the use of time series analytical techniques that require stationarity, such as ARIMA or Vector Autoregression models. Figure \ref{look3} displays the time series plot after the first differencing.

\subsection{Autocorrelation (ACF) and Partial Autocorrelation (PACF) Analysis}
After transforming the inflation rate data to achieve stationarity, the analysis employed Autocorrelation Function (ACF) and Partial Autocorrelation Function (PACF) plots \cite{peter2016introduction}. The ACF plot, which shows the correlation with lagged values, indicated successful differencing through a gradual decrease in autocorrelation. Conversely, the PACF plot displayed significant spikes at initial lags that quickly diminished, suggesting that only a few past values significantly influence the current value as shown in \textit{Supplementary Figure 1}. These analyses provided essential insights into the temporal structure of the inflation rate data, guiding the selection of autoregressive (AR) and moving average (MA) terms in subsequent modeling to enhance the robustness and predictive power of the econometric analysis.

\subsection{Autoregressive Integrated Moving Average, ARIMA (p, d, q)}

Building on insights from stationarity tests and autocorrelation analyses, the study applies the Autoregressive Integrated Moving Average (ARIMA) model, as formulated by Box and Jenkins \cite{box2015time}. The ARIMA modeling process consists of four key stages: identification, estimation, diagnostic checking, and forecasting. It is characterized by three components: autoregressive (AR), differencing (I), and moving average (MA), denoted as ARIMA (p, d, q). The parameter 'd' is set to 1 to achieve stationarity, while 'p' and 'q' are determined from the ACF and PACF plots. After specifying the model parameters, the ARIMA model is fitted to historical data by estimating the AR and MA coefficients. Diagnostic checks assess the model's adequacy by examining residuals for unaccounted autocorrelation. The fitted model is then used to forecast future inflation rates, providing valuable insights for economic planning and policymaking. The accuracy of these forecasts depends on the precise specification and fitting of the model, highlighting the importance of prior analyses.

\subsection{Implementation of Recurrent Neural Networks (RNN) and Long Short-term Memory
(LSTM)}

In this study, we utilized Recurrent Neural Networks (RNNs) \cite{medsker1999recurrent} for modeling time series data. RNNs are a class of neural networks particularly adept at processing sequential data, making them suitable for time series forecasting. Unlike traditional feedforward neural networks, RNNs have a memory that captures information about what has been calculated so far, essentially allowing them to have a sense of time. The architecture of RNNs, characterized by loops within the network, enables the capture of information in time-ordered sequences, making them particularly advantageous for predicting economic indicators such as inflation rates. We also employed Long Short-Term Memory (LSTM) networks \cite{hochreiter1997long}, a type of Recurrent Neural Network, for time series forecasting. LSTMs are particularly well-suited for time series data due to their ability to capture long-term dependencies, addressing the limitations of traditional RNNs like the vanishing gradient problem. The core of LSTM's architecture is the memory cell, which enables the network to retain information over extended sequences, crucial for accurate predictions in time series analysis.

\section{Results and Discussion}
\subsection{ARIMA Model}

To identify the most suitable ARIMA model for forecasting Ghana's inflation rates, the study utilized a stepwise search approach, recognized for its efficiency in model selection. This method systematically explores various model specifications to determine the optimal parameters (p, d, q) for the time series data. The Akaike Information Criterion (AIC) and Bayesian Information Criterion (BIC) served as guiding metrics, with lower values indicating a better fit while balancing model complexity. The stepwise search commenced with a set of initial models, adjusting parameters and comparing AIC values throughout the iterations. This process evaluated both the significance of each parameter and the model's overall performance, ensuring the selected ARIMA model effectively captured the inflation rate patterns without overfitting. The results of the stepwise search, detailed in Table \ref{tab}, identified ARIMA (1, 0, 1) as the optimal model for forecasting. This model comprises an autoregressive term of order 1 (AR(1)) and a moving average term of order 1 (MA(1)), with no differencing needed (d = 0), confirming that the series was already stationary. The choice of ARIMA (1, 0, 1) reflects its lower AIC score and its effectiveness in capturing both short-term dynamics and longer-term trends. 

\begin{table}[h!]
\centering
\caption{AIC values for various ARIMA models.}
\label{tab}
\vspace{0.2cm}
\begin{tabular}{|c|c|}
\hline
\textbf{Model Type} & \textbf{AIC Values} \\
\hline
ARIMA (2, 0, 2) & 265.871 \\
ARIMA (1, 0, 0) & 264.136 \\
ARIMA (0, 0, 1) & 265.142 \\
ARIMA (2, 0, 0) & 264.427 \\
ARIMA (1, 0, 1) & 262.855 \\
ARIMA (2, 0, 1) & 264.847 \\
ARIMA (1, 0, 2) & 264.846 \\
ARIMA (0, 0, 2) & 265.979 \\
\hline
\end{tabular}
\end{table}

\noindent Model diagnostics further validated the robustness of the chosen ARIMA model, as shown in \textit{Supplementary Figure 2} . Standardized residuals were inspected, and the absence of significant autocorrelation in the ACF plots suggested the model's adequacy. The Ljung-Box test reinforced these findings by confirming the residuals behaved as white noise, ensuring the model's reliability for forecasting inflation rates.


\subsection{Forecasting}
\begin{figure}[h!]
 \centering
  \includegraphics[width=10cm]{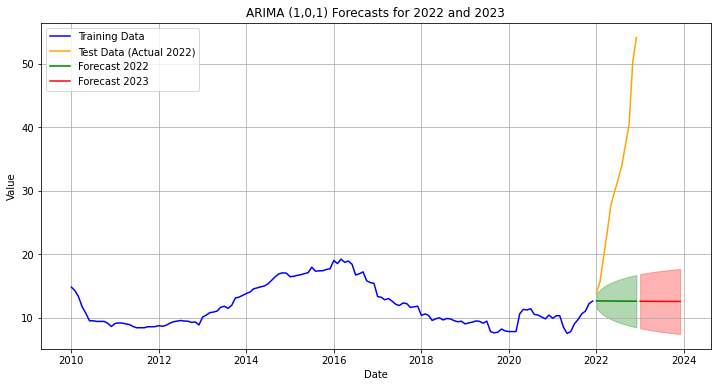}\\
  \caption{Forecasted Inflation rates for the Period 2022 - 2023 using ARIMA (1, 0, 1)}\label{look5}
\end{figure}

In the analysis shown in Figure \ref{look5}, a notable deviation between the forecasted and actual values is observed in the year 2022. 
This deviation can be attributed largely to the global impact of the COVID-19 pandemic, which began in early 2020 and continued to affect various sectors well into 2022. The pandemic brought about drastic changes in economic activities, consumer behavior, and governmental policies worldwide, leading to a high degree of volatility and unpredictability. The ARIMA (1, 0, 1) model used in our analysis, while robust under normal circumstances, could not foresee or account for such an extraordinary event. Standard time series forecasting models, including ARIMA, are typically built on the assumption of data continuity and trend stability derived from historical patterns. However, they lack the mechanism to incorporate sudden, external shocks like those induced by the COVID-19 pandemic. Consequently, the significant rise in the 2022 data underscores the limitations of conventional forecasting models in the face of global crises and highlights the
need for integrating adaptive, dynamic modeling approaches that can better handle such
unforeseen events.

\subsection{RNN and LSTM}
\begin{figure}[h!]
 \centering
  \includegraphics[width=11.8cm]{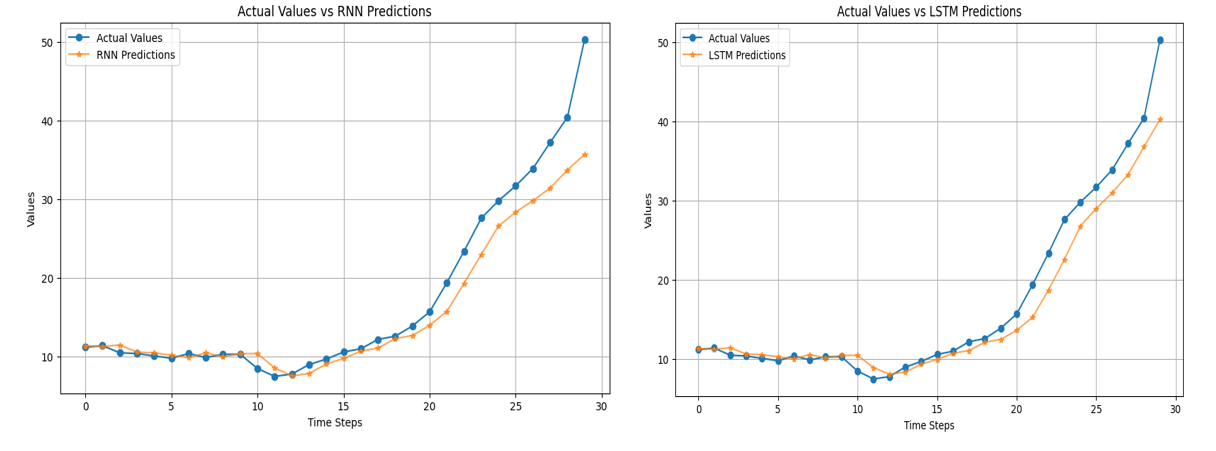}\\
  \caption{Left: Projections of Actual Inflation and RNN Predictions.
Right: Projections of Actual Inflation and LSTM Predictions.}\label{look4}
\end{figure}

This study employed RNN and LSTM models to analyze and forecast Ghana's inflation rates, beginning with data normalization for effective processing. The dataset was split into training and test sets, with 'look-back' periods used to incorporate lagged observations for future predictions. These periods are crucial for understanding inflation dynamics, as they determine how far back the model learns from past trends. Both models were configured with various layers and neurons to capture patterns in the inflation data. While RNNs were effective in identifying short-term trends, LSTMs excelled in recognizing longer-term patterns due to their ability to retain information over extended periods.
As illustrated in Figure \ref{look4}, both models performed well in tracking actual inflation rates during the initial and mid-time steps. However, as inflation values sharply increased, a divergence in performance emerged; the LSTM model closely aligned with actual values in later stages, while the RNN struggled to accurately reflect the magnitude of the inflation rise, likely due to limitations like the vanishing gradient problem. To substantiate these findings, analysis of quantitative performance metrics—such as Root Mean Square Error (RMSE), Mean Absolute Error (MAE), and Mean Absolute Percentage Error (MAPE)— provided an objective evaluation of each model's forecasting capabilities.

\subsubsection{Model Evaluation}
In the analysis, we evaluated the performance of the ARIMA model alongside RNN and LSTM networks for forecasting inflation rates, using three key metrics: RMSE, MAE, and MAPE. The results indicated that the LSTM model outperformed the RNN in predicting monthly inflation rates. Specifically, the LSTM achieved an RMSE of 2.80, MAE of 1.82, and MAPE of 8.82\%, demonstrating superior predictive capability, especially during periods of sharp inflation increases. In contrast, the RNN exhibited higher error rates, with an RMSE of 3.63, MAE of 2.14, and MAPE of 9.47\%, indicating a less accurate alignment with actual data. This quantitative analysis highlights the LSTM's strength in capturing complex temporal patterns, confirming its effectiveness for economic time series forecasting.

\section{Conclusion and Recommendations}

The study analyzed inflation trends in Ghana from 2010 to 2021, reserving the 2022 data for forecasting using advanced machine learning (ML) techniques, specifically Long Short-Term Memory (LSTM) and Recurrent Neural Network (RNN) models for time series forecasting. This approach revealed complex inflation patterns that traditional statistical methods could not uncover, identifying significant peaks and troughs influenced by various economic policies and external factors. The deviation in the 2022 forecast, likely due to the global impact of the COVID-19 pandemic, highlights the challenges of economic forecasting amid unprecedented shocks. Internal political and economic policies in Ghana, including fiscal and monetary decisions, significantly affect inflation rates. Changes in interest rates, taxation strategies, and public spending can lead to immediate impacts on inflation. Additionally, Ghana's reliance on commodities like cocoa and gold makes its economy sensitive to price fluctuations, which can swiftly influence inflation.
This study underscores the importance of adaptability in forecasting models and advocates for a data-driven approach in economic planning. Policymakers are encouraged to integrate ML forecasts into their decision-making processes to enhance the accuracy of economic policies. Future research should broaden the analytical horizon by incorporating various economic indicators, such as employment rates, GDP growth, and foreign exchange trends, to provide a more comprehensive view of economic health. Moreover, fostering interdisciplinary expertise between economists and data scientists can lead to more sophisticated models. The ongoing application and enhancement of ML models in economic analysis are crucial for ensuring Ghana’s economic stability and promoting sustainable development. As Ghana navigates the complexities of the global economy, the role of ML in shaping a resilient and thriving economy becomes increasingly vital.

\section*{Acknowledgments}
Our deepest gratitude goes to all who have supported and made invaluable contributions to this research in various ways.
\bibliographystyle{elsarticle-num}
\bibliography{reference}

\end{document}